\documentclass[useAMS,usenatbib, usegraphicx]{mn2e}
\usepackage{epsfig}
\usepackage{graphicx}
\usepackage{amsmath}
\usepackage{amssymb}
\usepackage{natbib}
\newcommand{\kms}{\ensuremath{{\rm km\,s}^{-1}}}
\newcommand{\mpc}{\ensuremath{{\rm Mpc}}}
\newcommand{\msun}{\ensuremath{M_{\odot}}}

\newcommand{\apj}{ApJ}

\newcommand{\mnras}{MNRAS}
\newcommand{\aj}{AJ}
\newcommand{\apjs}{ApJS}
\newcommand{\nat}{Nat}

\newcommand{\aap}{A\&A}

\begin{document}

\title[Overdensity of Shapley SC]{The Density Contrast of the Shapley 
Supercluster}

\author[Mu{\~n}oz \& Loeb]{Joseph A.\ Mu{\~n}oz
\thanks{E-mail:jamunoz@cfa.harvard.edu} and Abraham Loeb
\thanks{E-mail:aloeb@cfa.harvard.edu}\\Harvard-Smithsonian Center for
Astrophysics, 60 Garden St., MS 10, Cambridge, MA 02138, USA\\}

\maketitle

\begin{abstract}

We calculate the density contrast of the Shapley Supercluster (SSC) based
on the enhanced abundance of X-ray clusters in it using the extended
Press-Schechter formalism.  We derive a total SSC mass of
$M_{tot}=(4.4\pm0.44)\times10^{16}\,\msun$ within a sphere of $50\,\mpc$ 
centered at a distance of about $160\,\mpc$. The nonlinear fractional density
contrast of the sphere is $(1+\delta)=1.76\pm0.17$ relative to the mean
matter density in the Universe, but the contrast increases in the interior
of the SSC.  Including the cosmological constant, the SSC region is found
to be gravitationally unbound. The SSC contributes only a minor portion
($9.0\%\pm2.1\%$) of the peculiar velocity of the local group.

\end{abstract}


\section{Introduction}

The Shapley Supercluster (SSC) is one of the largest known structures in
the local Universe \citep{Fabian91, Quintana95, EFW97, Reisenegger00,
FSE05, Haines06, Proust06}.  Recent measurements based on the Tully-Fisher
relation give $11645\pm859\,\kms$ for the Hubble flow recession velocity of
A3558, one of the central clusters in the SSC \citep{Springob07}.  This
corresponds to a redshift of $z=0.0388$, a luminosity distance of $(166\pm13)\,\mpc$, 
and an angular diameter distance of $(154\pm11)\,\mpc$.
Despite the great distance of the SSC, it has recently been argued that the
SSC contributes significantly to the peculiar velocity of the local group
(LG) based on the enhanced abundance of X-ray clusters in it \citep{KE06}.

The contribution of the SSC to the LG peculiar velocity depends critically
on the mean matter density in the supercluster.  It is a non-trivial matter
to relate the enhanced abundance of massive X-ray clusters in the region to
its overdensity, since clusters are strongly biased with respect to the
underlying distribution of matter \citep{MW96,SMT01,Evrard02,Bahcall04}.
Because X-ray clusters correspond to rare density peaks, even mild
enhancements in the mean matter density can result in the formation of many
more clusters within the supercluster. Since this bias depends on the
cluster mass, the relationship between a sample containing objects in a
range of masses and the underlying matter cannot be described accurately by
a constant bias factor.

In this paper we study the dependence of the number of X-ray clusters in
the SSC on the underlying matter overdensity in this region.  We use the
hybrid model proposed by \citet{BL04} -- combining the extended
Press-Schechter (ePS) formalism \citep{Bond91} with the structure formation
prescription of \citet{ST99} (ST).  The model describes well published
results from numerical simulations (see discussion around Fig. 4 of
\citet{BL04}). Using the mass-dependent, nonlinear bias for X-ray clusters,
we calculate the matter overdensity in the SSC region from the number of
observed X-ray clusters there.

In \S \ref{method}, we describe how the \citet{BL04} hybrid model can be
used to calculate the mass in a large region enclosing collapsed objects.
We then specifically consider the SSC in \S \ref{clusters} and describe the
sample of x-ray clusters used in our calculation.  In \S \ref{SSC} we apply
our formalism and calculate the matter overdensity, cluster overdensity,
cluster bias, and total mass in the SSC region.  Given these results, we
then use the spherical collapse model in \S \ref{dynamics} to consider the
dynamics of the region. In \S \ref{pecvel}, we estimate the SSC contribution
to the peculiar velocity of the LG.  In \S \ref{radial}, we explore the
radial dependence of our results in the SSC and, finally, investigate the
robustness of our results in \S \ref{robust}.

We assume a flat, $\Lambda$CDM cosmology with the standard set of cosmological
parameters $\left( \Omega_{M,0}, \Omega_{\Lambda,0}, \Omega_{b,0}, h,
\sigma_8, \alpha, r\right) = \left( 0.279, 0.721, 0.0462, 0.701, 0.817,
0.960, 0.000\right)$ \citep{Komatsu08}.

\section{Method}\label{method}

According to the ePS prescription \citep{Bond91}, if the linear density
fluctuations in the universe are extrapolated to their values today and
smoothed on a comoving radius $R$, a point whose overdensity exceeds a
critical value of $\delta_{c}(z) \approx 1.68\,D(0)/D(z)$, belongs to a
collapsed object with a mass
$M=(4/3)\,\pi\,\rho_{crit,0}\,\Omega_{M,0}\,R^3$ if $R$ is the largest
scale for which the criterion is met. Here, $D(z)$ is the linear growth
factor at redshift $z$, $\rho_{crit,0}$ is the closure density of the
Universe, and $\Omega_{M,0}$ is the matter density parameter today.  The
critical value of the overdensity is also known as the {\it barrier}.  For
a Gaussian random field of initial density perturbations, as indicated by
measurements of cosmic microwave background (CMB)
\citep{Komatsu08}, the probability distribution of the extrapolated and
smoothed over-density, $\delta_R$, is also a Gaussian:
\begin{equation}\label{Pdel}
\frac{dP(\delta_R,S(R))}{d\delta_R}\,d\delta_R=\frac{1}{\sqrt{2\,\pi\,S(R)}}\,\exp\left(-\frac{\delta_R^2}{2\,S(R)}\right)\,d\delta_R,
\end{equation}
with zero mean and a variance given by:
\begin{equation}\label{var}
S\left(R\right)=\int_0^{k_{max}} \frac{dk}{2\,\pi^2}\,k^2\,P\left(k\right),
\end{equation}
where $P\left(k\right)$ is the linear power-spectrum of density
fluctuations today as a function of comoving wave-number $k$, and
$k_{max}=1/R$.  Since equation (\ref{var}) is a monotonically decreasing
function of $R$ (or $M$), the smoothing scale can be uniquely specified by
the variance of the over-density field smoothed on that scale.

If we define $f(\delta_c(z),S)\,dS$ to be the fraction of mass contained in
halos in the mass range $[M,M+dM]$ corresponding to $[S,S+dS]$ at redshift
$z$, then the comoving number density of collapsed objects in that mass
range is
\begin{equation}\label{mfdef}
\frac{dn}{dM}=\frac{\Omega_{M,0}\,\rho_{crit,0}}{M}\,\left|\frac{dS}{dM}\right|\,f(\delta_c(z),S).
\end{equation}
The ePS prescription, which uses a constant barrier independent of mass
with a value, $\delta_c(z)$, derived from the spherical collapse model,
gives
\begin{equation}\label{fPS}
f_{PS}(\delta_c(z),S)\,dS=\frac{1}{\sqrt{2\,\pi}}\,\frac{\nu}{S}\,\exp\left(-\frac{\nu^2}{2}\right)\,dS,
\end{equation}
where $\nu=\delta_c(z)/\sqrt{S}$ is the number of standard deviations a
density fluctuation on a scale $S$ must be above the mean to have crossed
the critical density threshold.  Incorporating a moving (i.e. scale
dependent) barrier generated from elliptical collapse with two free
parameters, the ST prescription gives \citep{ST99}
\begin{equation}\label{fST}
f_{ST}(\delta_c(z),S)\,dS=A\,\frac{\nu}{S}\,\frac{a}{\sqrt{2\,\pi}}\,\left[1+\frac{1}{(a\,\nu^2)^p}\right]\,\exp\left(-\frac{a\,\nu^2}{2}\right)\,dS
\end{equation}
and results in a mass function that better matches simulations.  The
best-fit values of the parameters are $a=0.75$ and $p=0.3$, while the
normalization factor is $A=0.322$ \citep{BL04}.

The unconditional mass function generated with $f(\delta_c(z),S)$
represents an average over all regions of space (or equivalently, over all
realizations of density Fourier mode amplitudes).  It assumes no prior
knowledge of the overdensity on a given scale.  If we fix the average
linear overdensity to be $\delta_L$ in a region smoothed on a particular
scale, $S_0$, we can generate a conditional mass function with a
conditional form of $f=f(\delta_c(z),S|\delta_L,S_0)$.  In ePS, this can be
done by substituting $\delta_c(z) \rightarrow \delta_c(z)-\delta_L$ and $S
\rightarrow S-S_0$ and gives results that agree well with simulations.  The
same substitution with $f_{ST}$, however, does not agree as well.  This is
because the new barrier height is scale dependent \citep{ZMF08}.

However, \citet{BL04} suggested using a hybrid model in which the conditional
mass function is generated from contributions by both $f_{PS}$ and $f_{ST}$
in regimes where they each fit best with simulations.  The resulting
collapsed mass fraction per variance interval is
\begin{equation}\label{fhybrid}
f_{hybrid}(\delta_c(z),S|\delta_L,S_0)=f_{ST}(\delta_c(z),S)\,\frac{f_{PS}(\delta_c(z),S|\delta_L,S_0)}{f_{PS}(\delta_c(z),S)}.
\end{equation}
This model describes well the numerical results from cosmological
simulations \citep{BL04}. Because $<\delta_L>=0$ in any region, the average
number of collapsed objects, $\bar{N}'$, with mass corresponding to $S$ in
a region corresponding to $S_0$ is generated by equation~(\ref{mfdef}) with
$f_{hybrid}(\delta_c(z),S|\delta_L=0,S_0)$.  We use this prescription to
map values of the overdensity in a region to the average number of
collapsed objects contained in it.  Variations between this conditional
average number of objects, $\bar{N}$ and the actually number, $N$, of those
counted in the region result only from Poisson fluctuations.  The cosmic
variance has been taken out by stipulating $\delta=\delta_L$ in $S=S_0$.

Given an integer number, $N$, of observed objects residing in a region
$S_0$, the differential probability distribution of values of $\bar{N}$
that have resulted in $N$ is,
\begin{equation}\label{PN}
\frac{dP(\bar{N}|N)}{d\bar{N}}=B\,P_{Poisson}(N,\bar{N})\,\frac{dP(\delta_L)}{d\delta_L}\,\frac{d\delta_L}{d\bar{N}},
\end{equation}
where
\begin{equation}
P_{Poisson}(k,\lambda)=\frac{\lambda^k\,e^{-\lambda}}{k!},
\end{equation}
is the Poisson distribution, $dP(\delta_L)/d\delta_L$ is the unconditional
distribution of overdensities in the region $S_0$ given by
equation~(\ref{Pdel}). The Jacobian $d\delta_L/d\bar{N}$ is derived from
equations~(\ref{mfdef}) and~(\ref{fhybrid}), and the coefficient $B$ is set
so as to normalize the integral of equation~(\ref{PN}) over $\bar{N}$ to
unity.

\begin{table*}
\caption{ X-ray Cluster Sample }
\begin{tabular}{lccccccccc}
\hline
Cluster & RA;Dec & $z$ & $r$ & $f_X$ & $k\,T$ & $M_{1625}$ & $M_{halo}$ & Refer. \\ 
& (J2000.0) & & & ($10^{-12}\,{\rm erg}$ & & & & \\ 
& (deg) & & ($\mpc$) & $\times\,{\rm cm^{-2}\,s^{-1}}$) & ($keV$) & ($10^{14}\,\msun$) & ($10^{14}\,\msun$) & \\ 
\hline
\hline
B6 & 194.795:-21.911 & --\,-- & 33 & 3.66 & 1.8 & 0.89 & 1.9 & A \\
A3548 & 198.379:-44.963 & --\,-- & 39 & 4.26 & 1.9 & 0.97 & 2.0 & A \\
B1 & 196.080:-17.001 & --\,-- & 44 & 7.53 & 2.5 & 1.5 & 3.0 & A \\
A721S & 196.513:-37.642 & 0.0490 & 23 & 7.76 & 2.5 & 1.5 & 3.0 & A \\
CIZA J1410.4-4246 & 212.619:-42.777 & 0.0490 & 41 & 9.35 & --\,-- & 1.7 & 3.5 & C \\
A1631 & 193.242:-15.379 & 0.0462 & 51 & 3.43 & 2.8 & 1.7 & 3.6 & A \\
A1736 & 201.758:-27.153 & 0.0453 & 17 & 27.54 & 3.0 & 1.9 & 4.0 & B \\
RXJ1332.2-3303 & 203.109:-33.812 & 0.0446 & 16 & 11.90 & 3.0 & 1.9 & 4.0 & A \\
3528S & 193.673:-29.231 & 0.0528 & 31 & 12.20 & 3.1 & 2.0 & 4.2 & A \\
A3530 & 193.917:-30.367 & 0.0537 & 33 & 9.24 & 3.2 & 2.1 & 4.4 & A \\
A3528N & 193.598:-29.010 & 0.0528 & 31 & 10.53 & 3.4 & 2.3 & 4.8 & A \\
RX J1252.5-3116 & 193.143:-31.266 & 0.0535 & 33 & 16.09 & 3.8 & 2.7 & 5.7 & A \\
SC 1327-312 & 202.514:-31.664 & 0.0495 & 6.6 & 12.25 & 3.8 & 2.7 & 5.7 & A \\
A3556 & 201.001:-31.656 & 0.0479 & 2.5 & 1.72 & 3.8 & 2.7 & 5.7 & A \\
A3528 & 193.640:-29.129 & 0.0528 & 31 & 24.32 & 4.0 & 3.0 & 6.2 & A \\
SC 1329-314 & 202.875:-31.812 & 0.0446 & 15 & 5.84 & 4.2 & 3.2 & 6.7 & A \\
A3562 & 203.446:-31.687 & 0.0490 & 5.6 & 29.16 & 4.3 & 3.3 & 6.9 & B \\
A3532 & 194.336:-30.375 & 0.0554 & 38 & 21.35 & 4.4 & 3.4 & 7.1 & A \\
A1644 & 194.332:-17.381 & 0.0473 & 45 & 4.45 & 4.6 & 3.7 & 7.6 & B \\
A3558 & 202.011:-31.493 & 0.0480 & 0.0 & 57.87 & 4.9 & 4.0 & 8.4 & B \\
A3571 & 206.860:-32.850 & 0.0391 & 40 & 110.9 & 6.8 & 6.6 & 13.7 & B \\
\hline
\end{tabular}
\begin{list}{}{}
\item{Col. (1): Source name.  Col. (2): RA and Dec (J2000).  Col. (3):
Redshift.  Col. (4): Distance from A3558 in $\mpc$.  Col. (5): $f_{\rm X}$
in the $0.1-2.4\ {\rm keV}$ energy band in units of $10^{-12}\ {\rm
ergs\,cm^{-2}\,s^{-1}}$.  Col. (6): Cluster average temperature in ${\rm
keV}$.  Col. (7): $M_{1625}$ calculated from equation~(\ref{M-T}) in units
of $\msun$.  Col. (8): $M_{halo} = 2.1\,M_{1625}$ in units of $\msun$.  We
conservatively estimate errors in $M_{halo}$ and $M_{1625}$ to be about
$30\%$.  Col. (9): Literature reference for the cluster temperature: (A)
\citet{FSE05}; (B) \citet{Vikhlinin08}; and (C) \citet{EMT02}.}
\end{list}
\label{sample}
\end{table*}

When applying this formalism to observations, there is an additional
complication in that overdensities and the sizes of regions are observed in
Eulerian coordinates which evolve as the region breaks away from the Hubble
flow, while the ePS and ST prescriptions rely on initial values in
Lagrangian coordinates.  Lagrangian sizes in the comoving frame do not
change over time.  As in \citet{ML08}, we use the spherical evolution model
to calculate the Lagrangian size, $R_L$, corresponding to the observed
Eulerian size, $R_E$, of a region containing the linear overdensity,
$\delta_L$ in a
${\Lambda}CDM$ universe.
The extent of the collapse depends on the magnitude of the overdensity.
The more overdense the region is, the larger it would have to be initially
in order for it to collapse to the same value of $R_E$.  Similarly, a lower
value of the overdensity would mean that the material inside $R_E$ came
from a relatively smaller Lagrangian size.  Integrating the equations of
motion results in a mapping between the comoving Eulerian size of a viewed
region and the comoving Lagrangian size of the region from where the same
material originated in the early universe.  For a fixed value of $R_E$,
there is a one-to-one relationship between $R_L$ and the value of
$\delta_L$ that collapses $R_L$ to $R_E$.  $R_L$ is then associated with
$S_0$.

Since matter shells in the spherical collapse model do not cross until
collapse, the amount of matter inside $R_L$ is the same as that inside
$R_E$.  The mass contained within the observed size $R_E$ is
\begin{equation}\label{mass}
M_{tot}(<R_E)=(4/3)\,\pi\,R_L^3\,\Omega_{M,0}\,\rho_{crit,0}\,(1+\delta_i),
\end{equation}
where the initial value of the overdensity is
$\delta_i=D(z_i)\,\delta_L/D(z_0))$, $z_i$ is the initial redshift before
the region begins to evolve nonlinearly, and $z_0$ is the observed redshift
of the region.  The nonlinear matter overdensity is then
$\delta=M/(V_{SSC}\,\Omega_{M,0}\,\rho_{crit,0})-1$, and the nonlinear bias
is $b=(N/\bar{N}'-1)/\delta$, where $V_{SSC}=(4/3)\,\pi\,R_E^3$ and
$\bar{N}'=n_{ST}\,V_{SSC}\,(R_L/R_E)^3$.  $n_{ST}$ is the unconditional, ST
mass function.

\section{The Cluster Sample}\label{clusters}

To calculate the mass and overdensity described in the previous section for
the SSC, we need to know the number of the collapsed objects in the SSC,
their minimum mass, and the size of the region in which they reside.  Since
the mass function given by ePS and similar prescriptions ignores
substructure, we must be careful to consider only the largest structures.
Fortunately, the largest collapsed halos are also the most sensitive
tracers of the mass function, being on its exponential tail.  We consider
the most massive, X-ray luminous clusters in the supercluster as tracers of
the most massive halos and make the assumption that each halo hosts one
such cluster.  This is a good approximation given how well the average ST halo mass function in the universe matches the cluster mass function \citep{Vikhlinin08}.  Our tendency to focus on only the rarest objects is
moderated by the need to reduce Poisson fluctuations on our sample.

To balance these requirements, we construct a sample of clusters in the SSC
whose host halos have masses above $M_{min}\sim1.75\times10^{14}\,\msun$.  The sample
consists primarily of clusters studied by \citet{FSE05}, but contains an
additional cluster from the {\it {Clusters in the Zone of Avoidance}}
(CIZA) sample \citep{EMT02}.  We have approximated the distance between
each of the clusters in our sample and the nominal center of the SSC, which
we have taken to be A3558.  To estimate these distances, we assume that
each cluster has the same peculiar velocity as A3558, measured by
\citet{Springob07}.  This is nearly equivalent to assuming no relative
peculiar velocity among the clusters.  As we will see in \S \ref{dynamics}
this approximation is adequate for our purposes.  Our entire sample extends
to a radius of roughly $R_E=50\,{\rm Mpc}$ around A3558.  At a distance of
$164\,\mpc$, $50\,\mpc$ perpendicular to the line-of-sight corresponds to
$\sim17.5\,{\rm degrees}$.  We assume that the clusters newly detected by
\citet{FSE05} have $z=0.048$.  We make the same assumption about $A3548$.
We exclude the object denoted as B8 by \citet{FSE05} because it is not a
``confirmed cluster".  Table~\ref{sample} gives our resulting sample of
$21$ clusters.  It includes a list of the clusters, positions, redshifts,
estimated distances to A3558, measurements of the X-ray flux and
temperature ($f_{X}$ and $T_{X}$, respectively), and the resulting
estimates of the host halo mass, $M_{1625}$ and $M_{halo}$, which are
defined below.  Our sample is similar to the $17$ clusters studied by
\cite{KE06} in the SSC in number and extent.

Here, $M_{1625}$ is the mass of the halo out to the radius within which the
matter density is $1625$ times the mean matter density\footnote{The notation $M_{500}$ used sometimes in the literature to refer to the halo mass out to the radius containing an average matter density equal to $500$ times the critical density (eg. \citet{Vikhlinin08}).  For our cosmology and at the redshift of the SSC, this notation corresponds to $M_{1625}$ used here and elsewhere in the literature (eg. \citet{HK03}) where the average matter density in the sphere is $1625$ times the mean matter density of the universe.}.  To calculate $M_{1625}$, we
make use of its relationship to X-ray temperature $T_X$ \citep{Vikhlinin08}:
\begin{equation}\label{M-T}
M_{1625} = (2.95\pm0.10)\times10^{14}\,\left(\frac{k_B\,T_X}{5\,{\rm keV}}\right)^{1.5}\,h^{-1}\,E(z)^{-1}\,\msun.
\end{equation}
For CIZA J1410.4-4246, we calculate $M_{1625}$ from the X-ray flux via the
relation \citep{Vikhlinin08}:
\begin{eqnarray}\label{M-L}
ln\,L_X & = & (47.392\pm0.085)+(1.61\pm0.14)\,ln\,M_{1625} \nonumber \\
& & + (1.850\pm0.42)\,ln\,E(z) - 0.39\,ln(h/0. 72) \nonumber \\
& & \pm (0.396\pm0.039), 
\end{eqnarray}
where the X-ray luminosity, $L_{X}=(4\,\pi\,d_L^2)\,f_X$, is in units of
${\rm ergs\,s^{-1}}$, $d_L$ is the luminosity distance, $M_{1625}$ is in
units of $\msun$, and $E(z)=\sqrt{(\Omega_{M,0}\,(1+z)^3+\Omega_{\Lambda})}$.

\begin{figure*}
\begin{center}
\includegraphics[width=\textwidth]{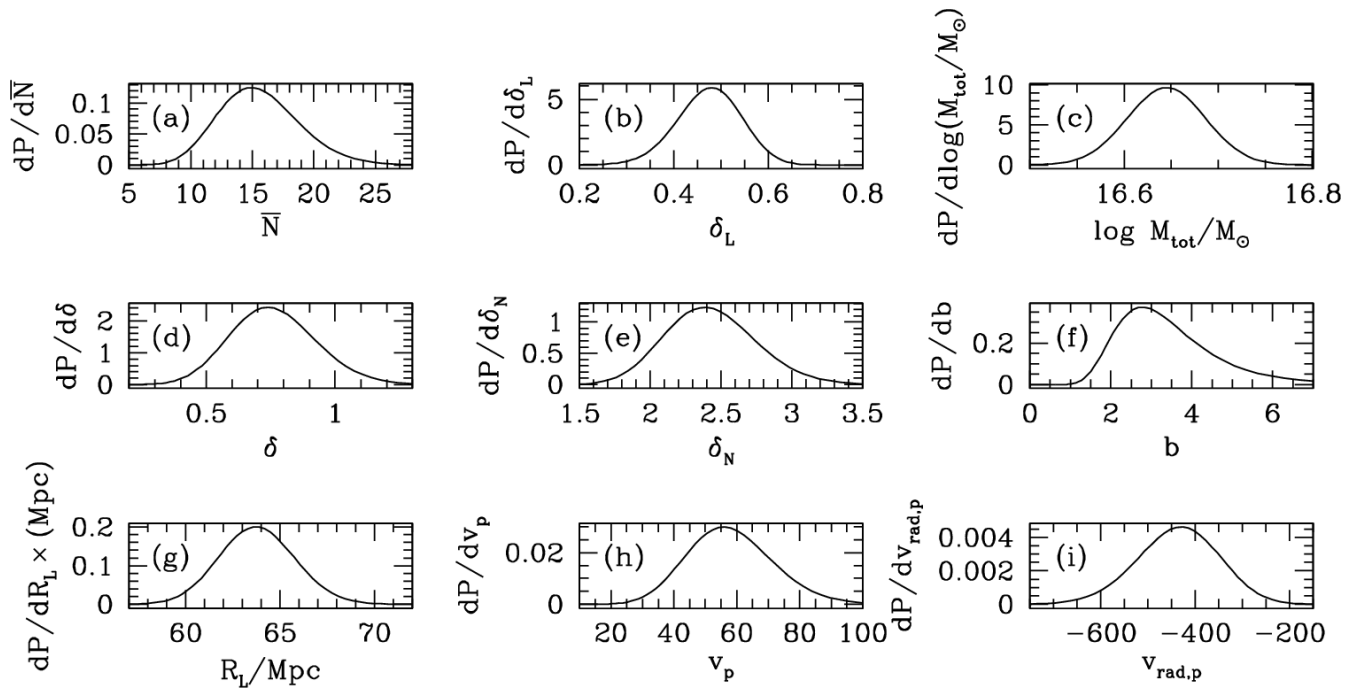}
\caption{\label{dist} Probability distributions of the expected number of
collapsed objects above $M_{min}$, the linear matter overdensity,
$\delta_L$, the total mass of the SSC, $M_{tot}$, the nonlinear matter
overdensity, $\delta$, the cluster overdensity, $\delta_N$, the cluster
bias, $b$, the Lagrangian radius of the SSC, $R_L$, the contribution of the
SSC to the peculiar velocity of the LG, $v_p$, and the peculiar radial
velocity of the outer edge of the SSC, $v_{rad,p}$.  $v_p$ and $v_{rad,p}$
are plotted in units of ${\rm km\,s^{-1}}$.  The distributions assume
$N=21$, $M_{min}=1.75\times10^{14}\,\msun$, and $R_E=51\,{\rm Mpc}$.  }
\end{center}
\end{figure*}

$M_{halo}$, on the other hand, is the mass for which ePS-like prescriptions
generate the mass function and corresponds roughly to the mass out to the
radius within which the matter density is $180$ times the mean density.
The relationship between various definitions of mass can easily be
calculated \citep{HK03}.  Values of $M_{1625}$ underestimate those of
$M_{halo}$ by $\sim52\%$ (i.e. $M_{halo} = 2.1\,M_{1625}$).  We estimate
errors in these masses to be about $30\%$, resulting primarily from error in
the temperature measurement and intrinsic scatter in the $M-T_X$ and
$M-L_X$ relations from equations~(\ref{M-T}) and~(\ref{M-L}).

Because the clusters under consideration are very massive and luminous, we
ignore the effect of the cluster selection function.  The flux of our
dimmest cluster is bright enough to be seen even in regions where ROSAT
sensitivity is slightly lower (i.e. in the $\sim 1.5\,{\rm deg}$ patch
centered at (RA:DEC)$=$(199.295:-34.393)).  However, absorption by the
galaxy could hide a few additional clusters.  On the other hand, because
the core of the SSC is clearly more overdense than the region out to
$50\,{\rm Mpc}$, our assumption that each cluster represents one halo may
not hold there.  Multiple clusters in the core may lie within a single halo
(or multiple halos in the process of merging).  Thus, the number of halos
may be somewhat different than the $21$ we assume.  Our analysis, however,
is independent of the individual masses of these halos and depends only on
the minimum halo mass of our sample and the total number of halos.  Because
the error in our mass measurements is only $30\,\%$, errors in the minimum
mass can be described by additional error in the number of clusters being
above a fixed threshold.  In addition, we expect the variation in $N$ due
to errors in $M_{halo}$ to be small.  If it turns out that all of our mass
estimates are systematically $30\,\%$ too high, then only two clusters
below the limit would have made it into our sample undeserved.  In \S
\ref{robust}, we show the dependence of our results on the exact value of
$N$.

\section{Shapley Overdensity}\label{SSC}

In Figure~\ref{dist}, we plot the distributions of several quantities
calculated through the prescription outlined in \S \ref{method} or derived
from the resulting quantities.  Panels (a)-(i) show results for the
expected number of collapsed objects above $M_{min}$, the linear matter
overdensity, $\delta_L$, the total mass of the SSC, $M_{tot}$, the
nonlinear matter overdensity, $\delta$, the cluster overdensity,
$\delta_N$, the cluster bias, $b$, the Lagrangian radius of the SSC, $R_L$,
the contribution of the SSC to the peculiar velocity of the LG, $v_p$, and
the peculiar radial velocity of the outer edge of the SSC, $v_{rad,p}$.

The distribution of $\bar{N}$ in panel (a) shows a mean clearly less than
our assumed value of $N=21$.  This is due to the
$(dP/d\delta_L)\,(d\delta_L/d\bar{N})$ factor in equation~(\ref{PN}).  If,
in the absence of information about $N$, each value of $\bar{N}$ were
equally likely, we would have $<\bar{N}>=N$.  However, including the true
prior probability distribution means that lower values of $\bar{N}$ are
more likely.

The linear overdensity can be interpreted given the value of
$\sigma=\sqrt{S}$ on the scale of the supercluster.  For $R=50\,{\rm Mpc}$
($64\,{\rm Mpc}$), $\sigma\approx0.23$ ($0.18$).  The mean linear
overdensity of $\sim0.47$ in panel (b) represents a $\sim2.0\sigma$
($\sim2.6\sigma$) fluctuation in the overdensity in the region.  As a
sanity check, we note that the probability of a linear density as high or
higher, assuming initial Gaussian fluctuations, is approximately $1/50$
($1/200$), and there are about $[200\,\mpc/50\,\mpc]^3\sim60$
($[200\,\mpc/60\,\mpc]^3\sim40$) such regions out to the distance of the
SSC.  So it is not unlikely that we do observe such a region within $200
{\rm Mpc}$, nor is it so unlikely that the supercluster remains one of the
most overdense regions out to that distance.

Our analysis gives $M_{tot}=(4.4\pm0.44)\times10^{16}\,\msun$ for the mass
of the SSC.  This is similar to the mass in the region estimated by
\citet{FSE05}, and only a bit lower than estimates by
\citet{Reisenegger00} and \citet{Proust06}.

Panels (d)-(f) show the probability distributions of the matter
overdensity, $\delta$, the cluster overdensity, $\delta_N$, and the cluster
bias, $b$.  The nonlinear overdensity distribution shows $\delta=0.76\pm0.17$ and $\delta_N=2.4\pm0.32$.  Larger values of $\delta$ correspond to lower
values of $\delta_N$ and $b$ since the number of clusters is fixed and so
more matter in the region gives closer agreement between the matter and
cluster densities.  The bias is $b=3.5\pm1.4$, however, the distribution is
not Gaussian and is skewed toward values higher than its peak at
$b\approx1.6$.  This peak value is just over $1-\sigma$ higher than the maximum of the range of $b=1.2$--$1.9$ that \citet{KE06} estimated (for $\Omega_{M,0}=0.279$) by assuming that their
calculation of the peculiar velocity of the LG induced by their sample is
equal to the true value.  We calculate only a $6\%$ chance that the bias falls
into the $b=1.2$--$1.9$ range.  For reference, we find that $b=1.5$
corresponds to $\delta\approx1.2$.  The discrepancy in the bias may result
from differences between the clusters associated with the SCC and the
entire \citet{KE06} sample.  The entire sample includes objects that are
less intrinsically luminous (i.e. less massive) than those in the SSC.

\section{Shapley Supercluster Dynamics}\label{dynamics}

While we used the spherical collapse model in previous sections to
calculate $R_L$ corresponding to $R_E$ for each value of $\delta_L$, we now
look at the results from the model in detail about the dynamics of the SSC
region.  We describe our analysis in \S \ref{collapse} and possible tests
for our results in \S \ref{testing}.

\subsection{The Spherical Collapse Model}\label{collapse}

The evolution of the
physical size $r$ of a region containing mass $M_{tot}$ is described by:
\begin{equation}\label{sphcol1}
\frac{d^2s}{da^2}=\left(\frac{1}{H\,a}\right)^2\,\frac{d^2s}{dt^2}-\left(\frac{1}{a}+\frac{1}{H}\frac{dH}{da}\right)\,\frac{ds}{da}
\nonumber
\end{equation}
\begin{equation}\label{sphcol}
\frac{d^2s}{dt^2}=H_i^2\,\left(-\frac{\Omega_{M,i}\,(1+\delta_i)}{2\,s^2}+\Omega_{\Lambda,i}\,s\right),
\end{equation}
where $M_{tot}$ is the mass in the region given by equation~(\ref{mass}),
$s=r/r_i$, $r$ is the evolving Eulerian size of the region, and where the evolution of the Hubble parameter is given by the
Friedmann equation,
\begin{equation}\label{hubble}
H=\frac{1}{a}\,\frac{da}{dt}=H_0\,\sqrt{\frac{\Omega_{M,0}}{a^3}+\Omega_{\Lambda,0}}\,.
\end{equation}
The evolution of the matter and dark energy density parameters
($\Omega_{M}$ and $\Omega_{\Lambda}$, respectively) are given by
\begin{equation}\label{omega}
\Omega_{M}=(1-\Omega_{\Lambda})=
{\Omega_{M,0}\over \Omega_{M,0}+\Omega_{\Lambda,0} a^3}.
\end{equation}
Subscripts ``0" refer to values today, and subscripts ``i" refer to initial
values at $a_i$.  We take $a_i=0.01$, set $s_i=1$ 
(i.e. $r=r_i$) and $(ds/da)_i=(1-\delta_i/3)\,(1/a_i)$ as our initial
conditions, and numerically integrate equation~(\ref{sphcol}) up to
$a_f=1/(1.0388)\approx0.962$, corresponding to the Hubble flow redshift of
A3558.

\begin{figure}
\begin{center}
\includegraphics[width=\columnwidth]{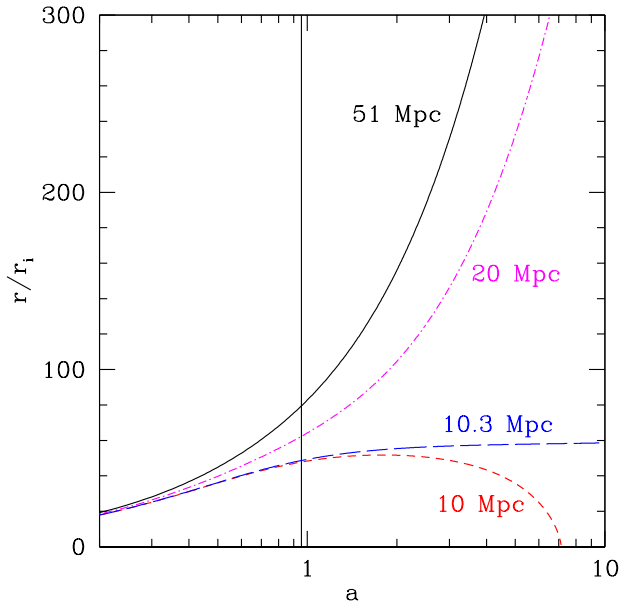}
\caption{\label{collpro} The evolution of spheres around the center of the SSC as a function of the scale factor, $a$, calculated from the spherical collapse model.  Results are shown for radii of $51\,\mpc$ (solid line), $20\,\mpc$ (dot-dashed line), $10.3\,\mpc$ (long-dashed line), and $10\,\mpc$ (short-dashed line).  Since $a=1$ today, values of $a>1$ refer to time in the future.  The vertical line denotes the redshift of the SSC.  The mean values of the overdensity within the radii shown have been adopted in plotting the trajectories.}
\end{center}
\end{figure}

In setting $(ds/da)_i=(1-\delta_i/3)\,(1/a_i)$,
as appropriate for a growing-mode perturbation, we have assumed that the
region is not initially expanding with the Hubble flow as in \citet{LH01}.
The $-\delta_i/3$ term accounts for the peculiar velocity of the region due
to its overdensity.  Since we expect the $\delta_i$ to be small, however,
the peculiar velocity term will also be small.  

The probability distribution for the Lagrangian radius, $R_L=r_i/\,a_i$, is
given in panel (g) of Figure~\ref{dist}.  The distribution gives
$R_L=(64\pm2.0)\,\mpc$, indicating that the region of the SSC deviates from
the Hubble flow and collapses by only $\sim20\%$.

The more interesting result from applying the spherical collapse model to
the SSC is that the region is still expanding at a radius of $50\,\mpc$.
While not expanding as fast as the Hubble flow, the SSC has not reached
turn-around.  Moreover, we find that, due to the repulsive effect of the
cosmological constant, the region will never reach turn-around despite its
overdensity but will continue to expand forever.  This can be seen in the trajectory 
of the $51\,\mpc$ region plotted in Figure~\ref{collpro}.  Trajectories have also been plotted for the interior regions of the SSC (see \S \ref{radial}).

We arrive at the
peculiar radial velocity of the region by subtracting the Hubble velocity
from the total expansion rate.  The distribution in the peculiar radial
velocity of the $50\,{\rm Mpc}$ shell, $v_{rad,p}$, is given in panel (h)
of Figure~\ref{dist} and gives $v_{rad,p}=(-437\pm87)\,\kms$.  The minus
sign indicates that the region is expanding somewhat slower than the
$\approx3600\,{\rm km\,s^{-1}}$ Hubble velocity at that radius.

\subsection{Observational Tests 
of the Model Predictions}\label{testing}

Our prediction for the outward velocity of the $50\,{\rm Mpc}$ shell could
potentially be tested with the Tully-Fisher relation or by Type Ia supernovae \citep{Masters06, Springob07, JRK07}.  By determining the
distance to an object of a known redshift, one may infer its peculiar
velocity by subtracting off the Hubble velocity at its inferred distance.  The quantity $v_{rad,p}$ is the radial peculiar velocity with respect to the center of the SSC.  Thus, the
peculiar velocity of the SSC itself must be accounted for when using the
Tully-Fisher relation to measure $v_{rad,p}$.  The latest results from
\citet{Springob07} provide a peculiar velocity for A3558, the central
cluster of the SSC, of $(2980\pm859)\,\kms$.  However, the error on the
peculiar velocity of A3558 is too large to constrain $v_{rad,p}$ reliably.
Below we explore alternative observational methods for future measurements
of the predicted peculiar velocities.

The peculiar velocity of an X-ray cluster can be directly inferred from its
contribution to the CMB through the kinetic Sunyaev-Zel'dovich (kSZ) effect
in which CMB photons scatter off free elections moving along with the bulk
velocity of the cluster.  The temperature fluctuation induced in this way
by a cluster is given by:
\begin{equation}\label{kSZ}
\frac{{\Delta}T}{T}=-\tau\,\frac{v}{c},
\end{equation}
where $\tau\approx\sigma_T\,n_{e}\,R_c$ is the optical depth for Thomson
scattering of the CMB by free electrons in the cluster, $\sigma_T$ is the
Thomson cross-section, $n_{e}$ is the number density of electrons in the
cluster, $R_c$ is its core radius, and $v$ is its peculiar velocity.  The
minus sign indicates that velocities away from us result in temperature
decrements.  Because A1644 is on the edge of the SSC (albeit the edge on
the sky as opposed to along the line of sight), we use it as a test case to
estimate ${{\Delta}T}/{T}$ due to $v_{rad,p}$ of a hypothetical cluster on
the $50\,\mpc$ shell and positioned along the line of sight to A3558.  The
radius of A1644 is $\sim0.7\,{\rm Mpc}$, while its X-ray luminosity is $L_X\approx10^{44}\,{\rm ergs\,s^{-1}}$ and
its temperature is about $6\times10^7$K \citep{FSE05}.  Accounting for the
X-ray luminosity through thermal bremsstrahlung we get
$n_{e}\approx5\times10^{-4}\,{\rm cm^{-3}}$.  If the hypothetical cluster
under consideration (whose properties are the same as A1644) is $50\,{\rm
Mpc}$ in front of A3558 directly along the line of sight so that
$v_{rad,p}$ is also along the line of sight, the CMB temperature
fluctuation due to $v_{rad,p}$ would be ${{\Delta}T}/{T}\approx-2\times10^{-6}$
with ${{\Delta}T}\approx-5\,{\rm \mu\,K}$.  Because the peculiar velocity
is radially inward toward the center of the SSC, if the cluster is between
us and A3558, then this velocity is away from us and results in a
temperature decrement.

The SSC as a whole also induces CMB signals through the kSZ effect from
to its own peculiar velocity and through the Rees-Sciama effect \citep{RS68}.  For an overdensity of $\delta=0.76$, the optical depth to electron
scattering through the SSC is $\tau=2.3\times10^{-5}$.  The resulting kSZ
signal due to the peculiar velocity of the structure is
${{\Delta}T}/{T}\approx-2.1\times10^{-7}$.  In the Rees-Sciama effect, the
evolution of the gravitational potential, $\phi=-G\,M_{tot}/R$, during the
photon crossing time of the system introduces additional temperature
fluctuations of order $\frac{{\Delta}T}{T}\sim \frac{\delta\phi}{c^2}$.  If
the potential becomes deeper (i.e. more negative) as a photon passes
through the region, the redshift experienced by the photon on its way out
of the potential well is greater than the blueshift it receives on its way
in and the CMB in that direction shows a temperature decrement.  However, 
between $z=0.051$ and $z=0.027$ (which spans the light crossing time for
the SSC of $\sim [100\,\mpc /c]$), the SSC grows from $50.47\,\mpc$ to
$51.54\,\mpc$.  For $M_{tot}=4.4\times 10^{16}\,\msun$, the change in the
potential results in ${{\Delta}T}/{T}\approx8.7\times10^{-7}$.  The kSZ and
Rees-Sciama effects due to the SSC as a whole act against each other.

The predicted signal can probed by a number of upcoming CMB observatories
such as {\it Planck} (http://planck.esa.int), the {\it Arcminute Cosmology
Bolometer Array Reciever} (http://cosmology.berkeley.edu/group/swlh/acbar),
the {\it Atacama Pathfinder Experiment}
(http://www.mpifr-bonn.mpg.de/div/mm/apex), the {\it South Pole Telescope}
(http://spt.uchicago.edu), and the {\it Atacama Cosmology Telescope} (ACT,
http://www.physics.princeton.edu/act).  With a temperature noise of
$\sim35\,{\rm \mu\,K}$ per pixel, WMAP is not sufficiently sensitive to
detect this effect.  On the other hand, ACT has a temperature noise of
${\delta}T\sim5\,{\rm \mu\,K}$ per pixel and a pixel size of $3\,{\rm
arcminutes^2}$.  At a distance of $(150-50=100)\,\mpc$, an A1644-like
cluster would take up $\sim600$ pixels, lowering the detectable level to
$\sim0.2\,{\rm \mu\,K}$.  Therefore, the kSZ signal due to this cluster is,
in principle, detectable with ACT at the $\sim26\,\sigma$ level.  Though
additional sources of error may make this detection more challenging
\citep{SKH05}, the major obstacle to measuring this signal is disentangling
it from the primary anisotropy that is at least an order-of-magnitude
larger and has an identical spectrum.  It is also important to note, that
some type of distance measure from something like the Tully-Fisher relation or from Type Ia supernovae is still necessary to determine which clusters are on the $50\,{\rm Mpc}$ shell.

\section{Contribution to Peculiar Velocity of the Local Group}\label{pecvel}

\begin{figure}
\begin{center}
\includegraphics[width=\columnwidth]{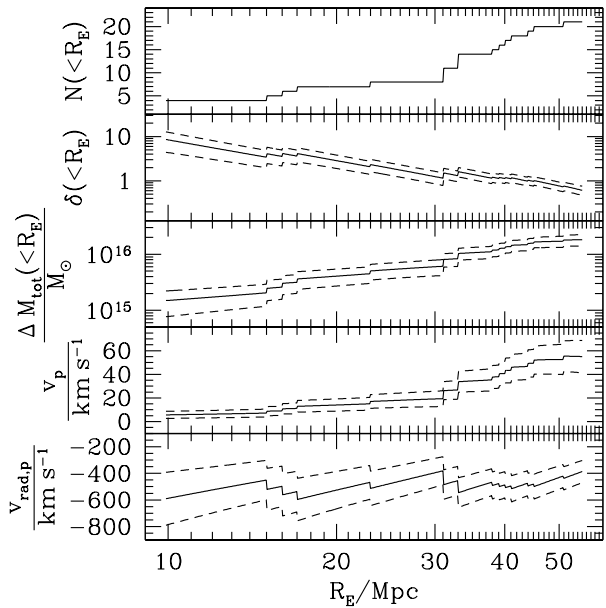}
\caption{\label{rprofile} The radial profile of the SSC.  The number of
observed clusters, $N$, the matter overdensity, $\delta$, and the amount of
excess mass, $\Delta\,M_{tot}$ within $R_E$ as functions of $R_E$ are
plotted in the top three panels.  The bottom two panels show the peculiar
velocity of the local group due to the overdensity contained within $R_E$
and the radial peculiar velocity of the shell with radius $R_E$ centered on
A3558.  Solid lines indicate mean values, while the dashed curves denote
$1-\sigma$ bounds due to Poisson fluctuations.  }
\end{center}
\end{figure}

Next we calaculate the contribution of the Shapley Suplercluster to the
peculiar velocity of the LG, $v_{p}$, given our result for the SSC mass.
The induced velocity in the linear regime\footnote{The expression for the
peculiar velocity given in equation~(\ref{v_p}) is only valid in the linear
regime.  However, since the SSC overdensity is modest, the formula gives a
good approximation to the true value.} is given by \citep{Peebles93}
\begin{equation}\label{v_p}
v_p=\frac{a\,f\,H}{4\,\pi}\,\int\,\frac{\vec{y}-\vec{x}}{\left|\vec{y}-\vec{x}\right|^3}\,\delta(\vec{y})\,d^3\vec{y},
\end{equation}
where $f=(a/D)\,(dD/da)\approx\Omega_M^{0.6}$ and $\vec{y}$ and $\vec{x}$
point from the center of the SSC to each SSC mass element and to the LG,
respectively.  Because we do not explore the mass distribution within the
SSC, we assume a constant value of the overdensity out to $50\,{\rm Mpc}$.
The distribution of values for $v_p$, reflecting the distribution in the
SSC overdensity, is plotted in panel (i) of Figure~\ref{dist}.  We find
$v_p=(55\pm13)\,\kms$, only $(9.0\pm2.1)\%$ of the $612\,\kms$ peculiar
velocity of the LG with respect to the CMB \citep{LN07}.  This is much less
than the value estimated by \citet{KE06} of $\sim30\%$.  Even $b=1.5$ and
the corresponding value of $\delta=1.2$ (see \S \ref{SSC}) produce only
$v_p\approx92\,{\rm km\,s^{-1}}$, but this is almost $3-\sigma$ from our mean value.  
Thus, we agree qualitatively with
\citet{Raychaudhury91}, \citet{Reisenegger00}, and \citet{LN07} that the
SSC does not contribute very significantly to the peculiar velocity of the LG.

\section{Radial Profile}\label{radial}

The number of clusters in our sample decreases sharply with radius from the
center of the SSC, and the clusters are not numerous enough to adequately
sample the entire volume of the SSC.  Nonetheless, we can obtain some
interesting estimates about the radial distribution of the quantities we
investigate by considering subsamples of clusters from Table~\ref{sample}
with distances from the center of the SSC within varying values of $R_E$.

Figure~\ref{rprofile} shows the radial profiles we calculate for the number
of observed clusters, the matter overdensity and the amount of excess mass
within $R_E$ as functions of $R_E$.  Also shown are the peculiar velocity
of the LG due to the overdensity contained within $R_E$ and the radial
peculiar velocity of the shell with radius $R_E$ centered on A3558.  The
plot of $N$ vs. $R_E$ depicts a step-like function where the value changes
in discrete steps as observed clusters from the Table~\ref{sample} become
included in the subsample within $R_E$.  The discreteness of the limited
sample is manifested in the sharp transitions exhibited in the radial
profiles of the other quantities in Figure~\ref{rprofile}.  The true
profiles are smooth curves that roughly follow those calculated and are
bounded by the $\pm 1-\sigma$ error brackets approximately $68\%$ of the
time despite the sharp transitions.  The uncertainty in the values of $r$,
the distance between each cluster and A3558, shown in Table~\ref{sample},
is not included in the errors plotted in the figure.

Since $\Delta\,M_{tot}=4\,\pi\,R_E^3\,\rho_M(z_{SSC})\,\delta$, the plots
of $\delta$ and $\Delta\,M_{tot}$ vs. $R_E$ are analogous, however, they
illustrate two different points.  The radial profile of $\delta$ shows a
roughly power-law rise in the overdensity as we move toward the center of
the SSC.  The supercluster becomes quite nonlinear in the region interior
to $R_E\approx40\,\mpc$.  The radial profile of $\Delta\,M_{tot}$, on the
other hand, demonstrates that the mass only rises significantly through
sharp jumps that correspond to the adding of additional clusters to the
subsample.  This can also be seen in the plot of $v_p$ vs. $R_E$ where the profile between jumps is extremely flat.  Moreover, while
the contribution to the peculiar velocity of the LG rises most rapidly
between $R_E\approx30\,\mpc$ and $R_E\approx45\,\mpc$, $v_p$ seems to have
leveled off by $R_E\approx50\,\mpc$.  

The evolution of $v_{rad,p}$ with $R_E$ is very mild with its mean varying by
only a factor of $\sim1.5$ over the full range of $R_E$ shown.  In addition,
the $1-\sigma$ errors on $v_{rad,p}$ are large, comparable to the range
over which $v_{rad,p}$ varies.  However, it is clear that the general trend is
toward increasingly negative values of $v_{rad,p}$ as we move toward the
center of the SSC.  This is what is expected from the spherical collapse
model.  In addition, by tracing the future of regions at different radii with the spherical collapse model (see \S \ref{collapse}), we determine that the SSC is only a closed system at $R_E\approx10.3$ given the mean of the density probability distribution and at $R_E\approx13$ if $1-\sigma$ fluctuations are considered.  The mass enclosed
within this radius is comparable to that of a rich X-ray cluster of $\sim
10^{15}M_\odot$.  The trajectories of spheres with $R_E=10\,\mpc$, $10.3\,\mpc$, $20\,\mpc$, and $51\,\mpc$ are plotted in Figure~\ref{collpro}.  The mean values of the overdensity for each of these radii have been adopted in plotting the trajectories.

\section{Robustness of Results}\label{robust}

In \S \ref{method}, we showed how the matter overdensity in a region can be
determined by fitting the mass function in an overdense region to the
cluster mass function in the region.  However, because of our limited
sample, we only fit the normalization of the mass function and relied on
the \citet{BL04} hybrid model to correctly fit the slope.  In this section,
we investigate the robustness of our results and their dependence on our
sample.  In \S \ref{Nrobust}, we consider the choice of $N$ obtained from
the sample, while in \S \ref{HMrobust}, we turn our attention to the slope
of the mass function and ask whether different mass bins within the sample
give consistent results.

\begin{figure}
\begin{center}
\includegraphics[width=\columnwidth]{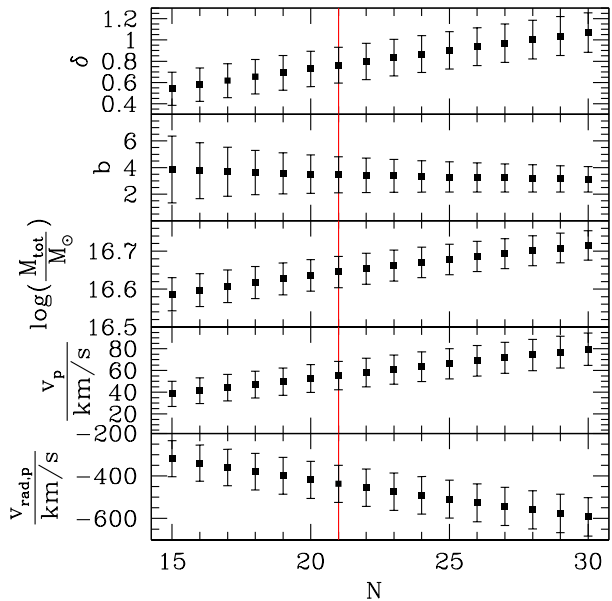}
\caption{\label{Neffect} The dependence of the distributions for the nonlinear
overdensity, $\delta$, the total mass of the SSC, $M$, the cluster bias,
$b$, the contribution of the SSC to the peculiar velocity of the LG, $v_p$,
and the radial peculiar velocity of the outer edge of the SSC, $v_{rad,p}$,
on the value of $N$.  The square points represent the mean values of the probability
distributions, while the error bars show the $\pm 1-\sigma$ spread.  The vertical line denotes $N=21$.  $M_{min}=1.75\times10^{14}\,\msun$ and $R_E=51\,\mpc$ are held fixed for all values of $N$.  }
\end{center}
\end{figure}

\begin{figure}
\begin{center}
\includegraphics[width=\columnwidth]{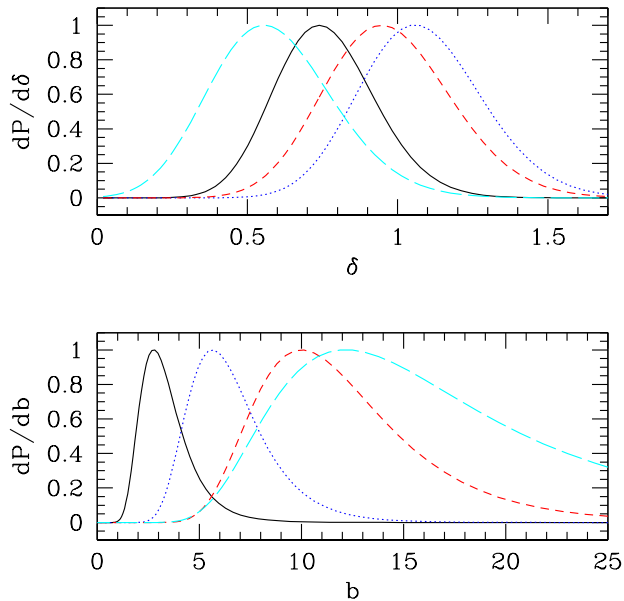}
\caption{\label{HMbins} The probability distributions of the matter
overdensity, $\delta$, (upper panel) and the cluster bias, $b$, (lower
panel) obtained with four different cluster subsamples from
Table~\ref{sample}.  The solid curve shows the results from the entire
sample of $21$ clusters with masses above $1.75\times10^{14}\,\msun$ shown in
Figure~\ref{Pdel}.  The dotted curve shows the results obtained using the
$17$ clusters with masses above $3.5\times10^{14}\,\msun$, while the $10$ clusters with masses above $5.5\times10^{14}\,\msun$ produced the results denoted by the dashed curve.  Finally, the
results denoted by the long-dashed curve were derived with only the $4$
clusters with host halo masses above $7.0\times10^{14}\,\msun$.  The size of
the region was fixed at $R_E=51\,\mpc$ for each distribution.}
\end{center}
\end{figure}

\subsection{Dependence on $N$}\label{Nrobust}

The number of collapsed halos, their minimum mass, and their extent are the
three key parameters that we extract from our cluster sample for use in our
structure formation analysis.  Though it is probably correct to identify
the number of X-ray luminous clusters with the number of collapsed halos,
the two may not be exactly equivalent.  In addition, there may be
undiscovered objects obscured by galactic absorption.  As argued in \S
\ref{clusters}, the error in the mass estimates of each halo from the
cluster X-ray properties can be expressed as an additional but small error
in the number of halos above a fixed threshold.  The same can be said for
the extent of the SSC, denoted $R_E$.  Small errors in the physical
position of a cluster with respect to the center of the SSC can be
represented as clusters being falsely included or excluded in our fixed
selection of $R_E$.  We thus explore the dependence of our results on the
assumed number of collapsed halos.

Figure~\ref{Neffect} shows the mean values of $\delta$, $b$, $M$,
$v_{rad,p}$, and $v_p$ from Figure~\ref{dist}, as well as the $1-\sigma$
errors for different choices of $N$.  As $N$ changes, $M_{min}$
and $R_E$ are held fixed at the values assumed throughout this
paper.  The results clearly do not change much as $N$ is varied over a
relatively wide range of values.

\subsection{Slope of the Mass Function}\label{HMrobust}

Checking the fit of the mass function explicitly is equivalent to asking
whether samples with higher mass thresholds give consistent results for the
overdensity in the SSC.  Since there are many more lower mass objects than
higher mass ones, significant deviations in the number of higher mass
objects may not affect the results obtained using all of the clusters.
However, consistency between multiple mass bins shows that the results do
not depend on the particular choices we made for our sample.

The probability distributions for the matter overdensity and the cluster
bias are shown in Figure~\ref{HMbins} calculated using each of four
different mass thresholds.  These distributions are not completely
independent because the lower mass thresholds include all of the objects at
higher mass.  However, since there is a steep drop-off in the number of
objects with mass, the distributions are dominated by objects near the mass
threshold.  All four mass thresholds give consistent results for the matter
overdensity in the SSC, with the distributions clustered around
$\delta\sim0.8$, slightly higher than the mean of the
$M_{min}=1.75\times10^{14}\,\msun$ distribution.  It is important to note that the
mass threshold does not correlate with mean overdensity.  This is just what
we would expect if there is no bias in the number of objects with mass
since, according to the model, a single value of the average linear matter
overdensity should describe the entire mass function.

On the other hand, we do expect a correlation between the bias and the mass
of the objects in the sample.  The figure clearly show this correlation as
the mean of each bias distribution increases with the mass threshold.  This
is a manifestation of the fact that more massive clusters are more strongly
clustered than less massive ones.

\section{Conclusions}\label{conclusions}

We have used the enhanced abundance of X-ray clusters to calculate the mass
in the SSC, based on the ePS-ST model \citep{BL04}.  We constructed a
sample of $21$ X-ray luminous clusters within a radius of $51\,{\rm Mpc}$
in halos with masses above $M_{min}=1.75\times10^{14}\,\msun$.  We then calculated
probability distributions for the matter density contrast of the SSC
region, the cluster bias for our sample, and the mass of the SSC.  We
demonstrated that even mild values of the overdensity in a region can
result in a significant over-abundance of massive clusters.  We found a
mass of $M=(4.4\pm0.44)\times10^{16}\,\msun$ for the SSC.  Our results are
in good agreement with previous results \citep{Reisenegger00, FSE05,
Proust06}, though we found that the cluster bias is probably higher than
that estimated by \cite{KE06}.

We then used the spherical collapse model to investigate the dynamics of
the SSC.  We found that the comoving size of the region has only collapsed
by about $20\%$ from its initial value.  Moreover, we concluded that the
SSC is not bound at a radius of $51\,{\rm Mpc}$, despite the significant
mass in the region.  The repulsive effect of the cosmological constant
provides the extra push against gravity that will keep the region from ever
collapsing.  The outer shell is moving radially away from the center with a
velocity only $(437\pm87)\,\kms$ slower than the $\approx3600\,{\rm
km\,s^{-1}}$ Hubble velocity at that radius.  This prediction could be tested with better Tully-Fisher data from SSC galaxies, or future CMB observations. 

We also calculated the contribution of the SSC to the peculiar velocity of
the LG, $v_p=(55\pm13)\,\kms$.  This value amounts to only
$(9.0\pm2.1)\%$ of the LG peculiar motion, much smaller than a recent
estimate by \citet{KE06} and more in agreement with \citet{Raychaudhury91},
\citet{Reisenegger00}, and \citet{LN07}.

Despite the uncertainty in investigating the interior of the SSC using
smaller subsamples of cluster, exploring the radial profiles of $\delta$,
$v_p$, and $v_{rad,p}$ indicated convergence, i.e.  that we have included
all of the mass in the region in excess of the universal average that
contributes to the peculiar velocity of the LG.  Moreover, while the
$50\,\mpc$ region as a whole is only mildly nonlinear, the interior of the
SSC becomes highly nonlinear.  The excess mass becomes enough to bind the 
sphere with radius only $\sim10\,\mpc$.

Finally, we showed that our results are robust to errors in our input
parameters and the mass threshold of our cluster sample.  Subsamples of
clusters with different mass thresholds give consistent results for the
overdensity in the region and demonstrate an expected increase in bias with
mass.

\section{Acknowledgements}

We would like to thank C. Jones, K. Masters, R. Narayan, and especially A. Vikhlinin for useful
discussions.  Thanks also to D. Kocevski for his list of X-ray clusters in the SSC.  JM acknowledges support from a National Science Foundation
Graduate Research Fellowship. This research was supported in part by
Harvard University funds.

\end{document}